\documentclass[aps,floatfix]{revtex4}
\usepackage{graphics,epsfig}
\usepackage{graphicx}

\begin{document}

\title{Models of Gravity in Five-dimensional Warped Product Spactimes}

\author{Sarbari Guha}
\affiliation{Department of Physics, St. Xavier's College (Autonomous), 30 Mother Teresa Sarani, Kolkata 700 016,India }
\author{Pinaki Bhattacharya}
\affiliation{Gopal Nagar High School, Singur 712409, West Bengal, India}

\begin{abstract}
In this paper, we have considered gravity in a 5-dimensional warped product space-time with a time-dependent warp factor, time-dependent extra dimension and a bulk cosmological constant. The braneworld is described by a spatially flat FRW-type metric. The five-dimensional field equations have been constructed and solved for various cases of the warp factor and the extra-dimensional scale factor, starting from very general type of the metric and then reducing down to specific cases. In the low energy regime, a stabilized bulk with constant curvature has been considered, from which the cosmological behavior of the corresponding observed universe has been interpreted. In the high energy regime, the bulk is assumed to be sourced by an ordinary massless scalar field and thereafter the type of the scalar field source and the different scale factors have been investigated.
\end{abstract}

\maketitle

\section{Introduction}

Theories of extra dimensions have become popular nowadays primarily due to their success in solving the hierarchy problem of particle physics \cite{aadd}\cite{add} and for the explanation that the post-inflationary epoch of our universe was preceded by the collision of D3-branes \cite{khoury}. These successes led us to the so-called 'braneworld scenario', where the ordinary standard model matter and non-gravitational fields are confined by some trapping mechanism to the 4-dimensional universe constituting the D3-branes (4-dimensional timelike hypersurfaces), embedded in a $(4 + n)$-dimensional 'bulk' (n being the number of extra dimensions). At low energies, gravity is confined to the brane along with particles, but at high energies gravity "leaks" into the higher-dimensional bulk, so that only a part of it is felt in the observed 4-dimensional universe. The strong constraints on the size of extra dimensions as in the Kaluza-Klein models, is therefore relaxed \cite{add}, lowering the fundamental Planck scale down to the TeV range. In these theories, the extra dimensions provide extra degrees of freedom on the brane, to compensate for any reduction in the number of degrees of freedom resulting out of some special features of the bulk.

Among the several higher-dimensional models developed over the years, the warped braneworld model of Randall and Sundrum with a single extra dimension \cite{rs1}\cite{rs2} has turned out to be popular for a number of reasons. In their model, matter fields were localized on a 4-dimensional hypersurface in a constant curvature five-dimensional bulk furnished with mirror symmetry. The exponential warp factor was responsible for "squeezing" the gravitational field closer to the 4-dimensional hypersurface. The metric was non-factorizable and the fifth dimension was infinite. The field equations on the corresponding 4-dimensional universe, were modified by the effect of the extra dimension. In their case, the warp factor was a function of the extra coordinate and the metric for the extra-dimensional coordinate was constant. However, both of these parameters can be functions of time and the extra coordinate. In the process, the solutions get very much complicated, and in many cases cannot be determined without the imposition of suitable constraints.

In this paper, we have considered RS-type braneworlds with the bulk in the form of a five-dimensional warped product space-time, having an exponential warp factor which depends both on time as well as on the extra-dimensional coordinate and a non-compact fifth dimension. By choosing the warp factor in such a form, we have taken into account a time-dependent process of localization of gravity as has already been considered by one of the authors \cite{GC}. It is then necessary to determine the modifications produced in the bulk as well as the consequences on the corresponding braneworld. The 5-dimensional bulk is assumed to have constant curvature. Although such a bulk has limited degrees of freedom, yet it is sufficient for us if the braneworld is of FRW-type, in which case no additional conditions are necessary to compensate for the reduced degrees of freedom. The advantage of assuming the braneworld to be of FRW-type is that it can be embedded in any constant curvature bulk including the de Sitter $dS_{5}$, anti de Sitter $AdS_{5}$ and the flat Minskowski manifold \cite{maia2}. Therefore, the braneworld considered by us is assumed to be defined by a flat FRW-type metric in the ordinary spatial dimension. However in their paper Guha and Chakraborty \cite{GC1} considered a simple type of warp factor, which was in the product form. In this paper, we have considered a more general kind of a time-dependent warp factor and have determined the nature of the solutions.

The plan of the paper is as follows. In Section II, the mathematical preliminaries have been considered. In Section III the relation between the extra-dimensional scale factor and the warp factor has been determined for the low energy regime and the corresponding field equations have been constructed. Due to the complicated nature of the field equations, the solutions cannot be obtained without imposing additional constraints. In Section IV, by imposing suitable constraints we have been able to arrive at the condition for a stabilized bulk with constant curvature, characterized by a negative cosmological constant. The corresponding solutions are obtained in Section V. Finally, in Section VI, we have considered the case of a bulk sourced by an ordinary massless scalar field in the high energy regime and have determined the solution for the scalar field and the relation between the warp factor and the scale factors. The conclusions have been summed up in Section VII.

\section{Mathematical preliminaries}

We consider a 5-dimensional theory described by an action of the form \cite{Nihei}
\begin{equation}\label{01}
S= -\frac{1}{2\kappa^{2}_{(5)}}\int d^{5}x \sqrt{\bar{g}}[\bar{R}+2\Lambda_{(5)}] + \int d^{4}x \sqrt{-\bar{g}}L_{m}
\end{equation}
where $g_{AB}$ is the 5-dimensional metric of signature (+ - - - -), $\Lambda_{(5)}$ is the bulk cosmological constant and $R$ is the 5-dimensional scalar curvature for this metric. The first term in (\ref{01}) corresponds to the Einstein-Hilbert action in 5-dimensions, the Lagrangian density $L_{m}$ represents all other contribution to the action which are not strictly gravitational, including the contribution of the matter fields localized on the brane, any interaction between the brane and the bulk and the brane itself. The constant $\kappa_{(5)}$ is related to the 5-dimensional Newton's constant $G_{(5)}$ and the 5-dimensional reduced Planck mass $M_{(5)}$ by the relation
\begin{equation}\label{02}
\kappa^{2}_{(5)}= 8\pi G_{(5)}= M^{-3}_{(5)}.
\end{equation}

With gravity being localized on the brane through the curvature of the bulk, the bulk cosmological constant depends on the curvature radius of the bulk manifold. For a bulk manifold in the form of a space of constant curvature $\bar{K}$, the magnitude of the bulk Riemann tensor is determined by the curvature radius '$l$' of the bulk manifold as follows:
\begin{equation}\label{03}
\bar{R}_{ABCD}= \bar{K}[\bar{g}_{AC}\bar{g}_{BD} - \bar{g}_{AD}\bar{g}_{BC}]= \frac{\epsilon}{l^{2}}[\bar{g}_{AC}\bar{g}_{BD} - \bar{g}_{AD}\bar{g}_{BC}].
\end{equation}
For $AdS_{5}$ geometry, $\epsilon=-1$, whereas for $dS_{5}$, $\epsilon=1$. The bulk cosmological constant is related to the curvature $\bar{K}$ by the relation $\Lambda_{(5)}= 6\bar{K}$.

The dynamics of the 5-dimensional space-time, including that of the 4-dimensional hypersurface representing the observed universe, is determined by the 5-dimensional field equations \cite{lrr}\cite{bdl}, which follows from (\ref{01}) as
\begin{equation}\label{04}
\bar{G}_{AB} = - \Lambda_{(5)}\bar{g}_{AB} + \kappa^{2}_{(5)}\bar{T}_{AB}
\end{equation}
where $\bar{G}_{AB}$ is the 5-dimensional Einstein tensor and $\bar{\bar{T}}_{AB}$ represents the 5-dimensional energy-momentum tensor.

We shall consider five-dimensional metrics with time-dependent warp factor as follows
\begin{equation}\label{05}
dS^2 =e^{2f(t,y)}\left(dt^2 - R^2(t)(dr^2 + r^2d\theta^2 + r^2sin(\theta)^2d\phi^2)\right) -A^2(t,y)dy^2
\end{equation}
where $y$ is the coordinate of the fifth dimension and $t$ denotes the conformal time. For the sake of simplicity, we have considered a metric having a flat spatial section. Here, $R(t)$ is the cosmological scale factor for the 4-dimensional hypersurface representing the observed universe, which is embedded in the 5-dimensional bulk. Further, $f$ is a smooth function, called the "warping function" and $e^{2f(t,y)}$ is the time-dependent warp factor. The observed universe is represented by the hypersurface $y=0$. The function $A(t,y)$ gives us a measure of the scale of the extra dimension at different times and at different locations  in the bulk. However, for a stabilized bulk, this function must settle down to a fixed value. For a conformally flat bulk metric \cite{GhoshKar}\cite{KT},  we can have the extra-dimensional scale factor in the form of a growing function of time, corresponding to a moving brane within a static bulk. A conformally flat bulk is a space-time of constant curvature and the resulting field equations are much simpler owing to the simplified geometry of the space-time.

The components of the energy-momentum tensor $\bar{T}$ for the bulk metric are given by

\begin{center}
$\bar{T}^{t}_{t}=\rho_{B},\qquad\qquad \bar{T}^{i}_{j}=-P_{B},\qquad\qquad \bar{T}^{y}_{y}=-P_{y}$.
\end{center}

\section{Field Equations in Five Dimensions with time-dependent warp factor and time-dependent extra dimension}

The non-vanishing components of the five-dimensional Einstein tensor for the space-time under consideration are

\begin{equation}\label{06}
\bar{G}^{t}_{t}=\frac{3}{e^{2f}} \left( \frac{\dot{R}^2}{R^2} + \frac{2\dot{R}\dot{f}}{R} + \dot{f}^2 + \frac{\dot{f}\dot{A}}{A} + \frac{\dot{R}}{R}\frac{\dot{A}}{A} \right) - \frac{3}{A^2}\left( 2f^{\prime 2} + f^{\prime\prime} -\frac{f^{\prime}A^{\prime}}{A} \right),
\end{equation}

\begin{equation}\label{07}
\bar{G}^{t}_{y}=-\frac{3}{e^{2f}} \left( (\dot{f})^{\prime} - \frac{\dot{A}}{A}f^{\prime} \right),
\end{equation}

\begin{equation}\label{07a}
\bar{G}^{y}_{t}= \frac{3}{A^2} \left( (\dot{f})^{\prime} - \frac{\dot{A}}{A}f^{\prime} \right),
\end{equation}

\begin{equation}\label{08}
\bar{G}^{y}_{y}= \frac{3}{e^{2f}} \left( \frac{\ddot{R}}{R} + \frac{\dot{R}^2}{R^2} + \frac{3\dot{R}\dot{f}}{R} + \dot{f}^2 + \ddot{f} \right)- \frac{6f^{\prime 2}}{A^2},
\end{equation}

and
\begin{equation}\label{09}
\bar{G}^{i}_{j}=\frac{1}{e^{2f}} \left( \frac{2\ddot{R}}{R} + \frac{\dot{R}^2}{R^2} + \frac{4\dot{R}\dot{f}}{R} + \dot{f}^2 + 2\ddot{f} + \frac{\dot{f}\dot{A}}{A} + \frac{2\dot{R}}{R}\frac{\dot{A}}{A} + \frac{\ddot{A}}{A} \right) - \frac{3}{A^2}\left( 2f^{\prime 2} + f^{\prime\prime} -\frac{f^{\prime}A^{\prime}}{A} \right).
\end{equation}

Above, a dot represents differentiation with respect to the conformal time $t$ and a prime stands for differentiation with respect to the fifth coordinate $y$. The conservation of the energy-momentum tensor $T^{a}_{b};a = 0$ leads us to the two equations
\begin{equation}\label{5a}
\dot{\rho}_{B} + 3 (\rho_{B} + P_{B} )( \dot{f} + \frac{\dot{R}}{R} ) + ( \rho_{B} + P_{y} )\frac{\dot{A}}{A} = 0,
\end{equation}
\begin{equation}\label{5b}
P^{\prime}_{y} + ( \rho_{B} - 3P_{B} + 4P_{y} )f^{\prime} = 0.
\end{equation}

In the brane-world scenario, the matter content of the observed 4-dimensional universe is assumed to be confined on a brane (which may have some thickness). However, the fifth component of the bulk energy-momentum tensor can be evenly distributed along the extra dimension. The other four components of the bulk energy-momentum tensor will arise out of the presence of a cosmological constant in the bulk. The presence of the non-zero $\bar{G}^{t}_{y}$ term indicates that matter or energy can escape from the 4-dimensional universe along the fifth dimension. However, the energy scale necessary to ensure such an access is few hundred GeV. The gravitational and gauge interactions unite at the electroweak scale and the observed weakness of gravity at long distances is due to the existence of large new spatial dimensions \cite{aadd}\cite{add}. General relativity fails to describe gravity at such high energies and has to be replaced by a quantum theory of gravity \cite{lrr}. At such scales of energy, gravity "leaks" into the higher-dimensional bulk and is propagated therein via the equivalence principle, so that only a part of it is felt in our 4-dimensions.

At low energies, gravity is localized at the brane along with particles and general relativity is recovered. In the five-dimensional theory proposed by Randall and Sundrum \cite{rs1}\cite{rs2}, the hierarchy between the fundamental five-dimensional Planck scale and the compactification scale was only of order 10 and the excitation scale was of the order of a TeV. To prevent matter or energy flowing out of the brane along the fifth dimension, we require that $\bar{T}^{t}_{y}=0$, which implies that $\bar{G}^{t}_{y}=0$ and hence we obtain
\begin{equation}\label{10}
\dot{f}^{\prime}=\frac{\dot{A}}{A}f^{\prime}.
\end{equation}
Assuming that both $A$, $f$ and their first order derivatives are continuous, (\ref{10}) can be easily integrated to give the result
\begin{equation}\label{11}
A(t,y)=\chi(y)f^{\prime}(t,y).
\end{equation}
Thus, the scale of the extra dimension at a given instant of conformal time $t$, depends on the way the warping function varies along the extra dimension at that instant and will be different at different locations in the bulk. Consequently it is expected that the manner in which gravity is localized on the 4-dimensional hypersurfaces (which is monitored by the exponential warp factor) at different locations in the bulk, will determine the scale of the extra dimension at those locations. However, at some other instant, the variation of the warping function along the extra dimension may be different and so the extra-dimensional scale factor at a specific location will also change.

The Einstein field equations for the above metric therefore reduce to the form
\begin{equation}\label{12}
\rho_{B} =\frac{3}{e^{2f}} \left( \frac{\dot{R}}{R} + \dot{f} \right) \left( \frac{\dot{R}}{R} + \dot{f} + \frac{\dot{f}^{\prime}}{f^{\prime}} \right) - \frac{3}{\chi^{2}f^{\prime}}\left( 2f^{\prime} - \frac{\chi^{\prime}}{\chi} \right),
\end{equation}
\begin{equation}\label{13}
-P_{B} = \frac{1}{e^{2f}} \left( 2 \left( \frac{\ddot{R}}{R} + \frac{\dot{R}\dot{f}}{R} + \ddot{f} \right) + \left( \frac{\dot{R}}{R} + \dot{f} \right)^{2} + \frac{\dot{f}^{\prime}}{f^{\prime}} \left( \dot{f} + \frac{2\dot{R}}{R} + \frac{\dot{f}^{\prime}}{f^{\prime}} - \frac{f^{\prime \prime}}{f^{\prime}} \right) + \frac{\ddot{f}^{\prime}}{f^{\prime}} \right) - \frac{3}{\chi^{2}f^{\prime}}\left( 2f^{\prime} - \frac{\chi^{\prime}}{\chi} \right).
\end{equation}
and
\begin{eqnarray*}
-P_{y}= \frac{3}{e^{2f}} \left( \frac{\ddot{R}}{R} + \frac{\dot{R}^2}{R^2} + \frac{3\dot{R}\dot{f}}{R} + \dot{f}^2 + \ddot{f} \right) - \frac{6}{\chi^{2}}
\end{eqnarray*}
\begin{equation}\label{14}
=\frac{3}{e^{2f}} \left( \frac{\ddot{R}}{R} + \left( \frac{\dot{R}}{R} + \dot{f} \right)^{2} + \frac{\dot{R}\dot{f}}{R} + \ddot{f} \right) - \frac{6}{\chi^{2}}
\end{equation}

Thus we have six unknowns ($f$, $R$, $\chi$, $\rho_{B}$, $P_{B}$ and $P_{y}$) and three independent equations. We therefore need to impose additional constraints to solve the field equations.

\section{The case of a stabilized bulk of constant curvature in the low energy regime}

The function $A(t,y)$ gives us a measure of the scale of the extra dimension at different times and different locations $y$ within the bulk. We assume that the 4-dimensional hypersurface representing the observed universe is located at the position $y=0$ in the bulk. For a stabilized bulk, the extra dimensional scale factor should not vary with time i.e. we must have $\dot{A}=0$. In that case, we can assume the extra-dimensional scale factor to depend only on $y$. Thus the components of the five-dimensional Einstein tensor for the space-time are now

\begin{equation}\label{15}
\bar{G}^{t}_{t}=\frac{3}{e^{2f}} \left( \frac{\dot{R}^2}{R^2} + \frac{2\dot{R}\dot{f}}{R} + \dot{f}^2 \right) - \frac{3}{A^2}\left( 2f^{\prime 2} + f^{\prime\prime} -\frac{f^{\prime}A^{\prime}}{A} \right),
\end{equation}

\begin{equation}\label{16}
\bar{G}^{t}_{y}=-\frac{3(\dot{f})^{\prime}}{e^{2f}},
\end{equation}

\begin{equation}\label{16a}
\bar{G}^{y}_{t}= \frac{3(\dot{f})^{\prime}}{A^2},
\end{equation}

\begin{equation}\label{17}
\bar{G}^{y}_{y}= \frac{3}{e^{2f}} \left( \frac{\ddot{R}}{R} + \frac{\dot{R}^2}{R^2} + \frac{3\dot{R}\dot{f}}{R} + \dot{f}^2 + \ddot{f} \right)- \frac{6f^{\prime 2}}{A^2},
\end{equation}

and
\begin{equation}\label{18}
\bar{G}^{i}_{j}=\frac{1}{e^{2f}} \left( \frac{2\ddot{R}}{R} + \frac{\dot{R}^2}{R^2} + \frac{4\dot{R}\dot{f}}{R} + \dot{f}^2 + 2\ddot{f} \right) - \frac{3}{A^2}\left( 2f^{\prime 2} + f^{\prime\prime} -\frac{f^{\prime}A^{\prime}}{A} \right).
\end{equation}
To prevent matter or energy from flowing out of the brane we impose the the restriction that $f$ must not vary with the conformal time. Hence the field equations get simplified to the form
\begin{equation}\label{19}
\rho_{B} = \frac{3}{e^{2f}} \frac{\dot{R}^{2}}{R^{2}} - \frac{3}{A^2}\left( 2f^{\prime 2} + f^{\prime\prime} -\frac{f^{\prime}A^{\prime}}{A} \right),
\end{equation}
\begin{equation}\label{20}
-P_{B} = \frac{1}{e^{2f}} \left( \frac{2\ddot{R}}{R} + \frac{\dot{R^{2}}}{R^{2}} \right) - \frac{3}{A^2}\left( 2f^{\prime 2} + f^{\prime\prime} -\frac{f^{\prime}A^{\prime}}{A} \right)
\end{equation}
and
\begin{equation}\label{21}
-P_{y} = \frac{3}{e^{2f}} \left( \frac{\ddot{R}}{R} + \frac{\dot{R}^{2}}{R^{2}} \right) - \frac{6f^{\prime 2}}{A^2}.
\end{equation}

In order that the specific bulk represents a space of constant curvature, the condition (\ref{03}) reduces to the following equality:
\begin{equation}\label{22a}
- \frac{\ddot{R}}{Re^{2f}} + \frac{f^{\prime 2}}{A^2} = - \frac{\dot{R}^{2}}{R^{2}e^{2f}} + \frac{f^{\prime 2}}{A^2} = \frac{1}{A^2}\left( f^{\prime\prime} - \frac{f^{\prime}A^{\prime}}{A} + f^{\prime 2} \right) = constant.
\end{equation}
This will be satisfied, for this constant curvature bulk if the following conditions hold
\begin{equation}\label{22b}
\frac{\ddot{R}}{R} = \frac{\dot{R}^2}{R^2},
\end{equation}
and
\begin{equation}\label{22c}
f^{\prime\prime} - \frac{f^{\prime}A^{\prime}}{A} = 0.
\end{equation}

The non-zero components of the Weyl tensor for the bulk metric are of the form
\begin{equation}\label{23}
C_{ABAB}=\pm K\frac{g_{AA}g_{BB}}{g_{tt}}\times\Phi
\end{equation}
where
\begin{equation}\label{24}
\Phi = \left(  \frac{\ddot{R}}{R} - \frac{\dot{R}^2}{R^2} \right).
\end{equation}

In view of (\ref{22b}), this means that the Weyl tensor for the bulk space-time vanishes \cite{Kramer}\cite{HE}, as is expected. Consequently, the bulk satisfies the condition $\rho_{B} = -P_{B}$. For such a space-time, the deceleration parameter for the observed universe is given by the unique value
\begin{equation}\label{25}
q = -1.
\end{equation}
We can therefore conclude that as a result of the constraints imposed on the bulk space-time, the observed universe is found to be in a state of uniform acceleration. This is expected during the late-time evolution of the observed universe. This means that the constraints which were imposed, reduced the space-time geometry to a form which is suitable for modeling the universe at an epoch of uniform acceleration.

\subsection{Solutions}

Let us now assume that $P_{B} = P_{y}$, i.e. the pressure in the bulk is isotropic and consequently the bulk is characterized by the energy density contributed by a negative bulk cosmological constant. When we couple this condition with the condition for the constant curvature bulk, following some straightforward calculation, we get the solution

\begin{equation}\label{26}
R(t) = sinh(\alpha t) + cosh(\alpha t)
\end{equation}
where $\alpha$ is a constant of integration and
\begin{equation}\label{27}
\frac{f^{\prime}}{A} = \xi(y),
\end{equation}
where both $f$ and $A$ are now functions of $y$ only. In this case, the warp factor depends only on the extra-dimensional coordinate and hence do not depend on time. Thus the the process of localization of gravity on a particular 4-dimensional hypersurface will only depend on the position of the hypersurface in the bulk and the manner in which the warping function $f$ varies along the extra dimension at that location. It will not depend on time.

\section{The case of a bulk characterized by a massless scalar field in the high energy regime}

Let us now assume that the bulk energy-momentum tensor is sourced by an ordinary, massless scalar field. To simplify the analysis, we consider the bulk metric to be described by
\begin{equation}\label{28}
dS^2 =e^{2f(y)}\left(dt^2 - R^2(t)(dr^2 + r^2d\theta^2 + r^2sin(\theta)^2d\phi^2)\right) -A^2(t,y)dy^2.
\end{equation}
For such a bulk, the non-zero components of the Einstein tensor are given by
\begin{equation}\label{29}
\bar{G}^{t}_{t}=\frac{3}{e^{2f}} \left( \frac{\dot{R}^2}{R^2} + \frac{\dot{R}}{R}\frac{\dot{A}}{A} \right) - \frac{3}{A^2}\left( 2f^{\prime 2} + f^{\prime\prime} -\frac{f^{\prime}A^{\prime}}{A} \right),
\end{equation}
\begin{equation}\label{29}
\bar{G}^{t}_{y}=\frac{3}{e^{2f}} \frac{\dot{A}}{A}f^{\prime},
\end{equation}
\begin{equation}\label{30}
\bar{G}^{y}_{t}= - \frac{3 \dot{A} f^{\prime}}{A^3},
\end{equation}
\begin{equation}\label{31}
\bar{G}^{i}_{j}=\frac{1}{e^{2f}} \left( \frac{2\ddot{R}}{R} + \frac{\dot{R}^2}{R^2} + \frac{2\dot{R}}{R}\frac{\dot{A}}{A} + \frac{\ddot{A}}{A} \right) - \frac{3}{A^2}\left( 2f^{\prime 2} + f^{\prime\prime} -\frac{f^{\prime}A^{\prime}}{A} \right),
\end{equation}
and
\begin{equation}\label{32}
\bar{G}^{y}_{y}= \frac{3}{e^{2f}} \left( \frac{\ddot{R}}{R} + \frac{\dot{R}^2}{R^2} \right)- \frac{6f^{\prime 2}}{A^2}.
\end{equation}
For this case we consider the possibility of the particles to escape into the extra dimension, so that the corresponding energy scale is in the GeV range. The general form of the bulk energy-momentum tensor for scalar field source in the bulk is given by
\begin{equation}\label{33}
\bar{T}^{scalar}_{IJ} = \partial_{I}\phi \partial_{J}\phi - \frac{1}{2}g_{IJ}\partial_{K}\phi \partial^{K}\phi
\end{equation}

The components of the bulk energy-momentum tensor for the given metric are
\begin{equation}\label{34}
\bar{T}^{t}_{t} = \frac{1}{2} \dot{\phi}^{2} e^{-2f} + \frac{\phi ^{\prime 2}}{2 A^2} = - \bar{T}^{y}_{y},
\end{equation}
\begin{eqnarray*}
\bar{T}^{i}_{i} = \frac{\phi ^{\prime 2}}{2 A^2} - \frac{1}{2} \dot{\phi}^{2} e^{-2f}
\end{eqnarray*}
and
\begin{eqnarray*}
\bar{T}^{t}_{y} = \dot{\phi} \phi^{\prime}e^{-2f}.
\end{eqnarray*}
To obtain the solutions for the scalar field, let us assume that
\begin{equation}\label{35}
\phi(t,y) = \phi_{1}(t) + \phi_{2}(y).
\end{equation}
From the $^{t}_{y}$-component of the bulk energy momentum tensor, we find that
\begin{center}
$\dot{\phi}_{1} = k \frac{\dot{A}}{A}$ \qquad \qquad \qquad and \qquad \qquad \qquad $\phi_{2}^{\prime} = \frac{3}{k}f^{\prime}$.
\end{center}
Consequently, the solution for the bulk scalar field turns out to be
\begin{equation}\label{36}
\phi(t,y) = k ln A + \frac{3}{k}f.
\end{equation}
If we now consider the geometry for a vanishing Weyl tensor, then we find that
\begin{equation}\label{37}
\bar{G}^{i}_{j}=\frac{3}{e^{2f}} \left( \frac{\ddot{R}}{R} + \frac{\dot{R}}{R}\frac{\dot{A}}{A} \right) - \frac{3}{A^2}\left( 2f^{\prime 2} + f^{\prime\prime} -\frac{f^{\prime}A^{\prime}}{A} \right).
\end{equation}
In this case, to have an isotropic bulk, we must have
\begin{equation}\label{38}
A(t,y)=af(y)R(t),
\end{equation}
where $a$ is a proportionality constant and
\begin{equation}\label{39}
f^{\prime}(y) = \eta(y)A(t,y).
\end{equation}

\section{Conclusions}
In this paper, we have consider five-dimensional warp product space-times having time-dependent warp factor and a non-compact fifth dimension. The warp factor reflects the confining role of the bulk cosmological constant to localize gravity at the brane through the curvature of the bulk. This process of localization may include some time-dependence. Hence we have considered a warp factor which depends both on time as well as on the extra coordinate. The extra dimensional scale factor is also a function of time and of the extra coordinate. Consequently, the field equations became very complicated. Therefore in their previous paper \cite{GC1}, one of the authors considered a simplified analysis, where the solutions were sought in the low energy regime (within the TeV scale), so that the geometry represented a stabilized bulk. Further, the bulk was assumed to be a space of constant curvature and the warp factor in the product form: a function of time and a function of the extra-dimensional coordinate $y$.

In this paper, we have considered a general form of a time-dependent warp factor, given by $e^{2f(t,y)}$ and have obtained the relation between the warp factor and the extra-dimensional scale factor in the low energy regime, when gravity and particles remain confined to the observed universe. However, to solve the field equations, we had to impose certain constraints. Namely, we considered a stabilized bulk in the low energy regime and having a constant curvature. In that case, the observed universe is found to be in a state of uniform acceleration and the energy density of the bulk is contributed by a negative bulk cosmological constant. The solution for the cosmological scale factor has been determined and the relation between the warp factor and the extra-dimensional scale factor has also been obtained. Finally, we considered a bulk sourced by a massless scalar field in the high energy regime and determined the solution for the scalar field. When such a bulk is of constant curvature and is characterized by isotropic pressure, we could obtain the relation between the extra-dimensional scale factor, the warp factor and the cosmological scale factor.

\section*{Acknowledgments}
A portion of this work was done in IUCAA, India under the associateship programme. SG gratefully acknowledges the warm hospitality and the facilities of work at IUCAA.

\end{document}